\documentclass[11pt,prd,superscriptaddress,nofootinbib]{revtex4}
\usepackage{graphicx}
\usepackage[usenames,dvipsnames]{xcolor}

\usepackage{amssymb,amsmath,amsfonts}

\begin{document}
\begin{titlepage}
	
	%---------------- preprint number ---------------
	\hfill\parbox{5cm} { }
	
	\vspace{25mm}
	
	\begin{center}
		%------------------------ title ------------------------
		{\Large \bf Isospin symmetry of $\omega$ meson in the  soft-wall model of holographic QCD at finite temperature}
		
		%---------------- authors and addresses ----------------
		\vskip 1. cm
		
		{Narmin Nasibova \footnote{Corresponding author:\\n.nesibli88@gmail.com}}
		\vskip 0.5cm

		 Institute of Physics, Ministry of Science and Education of Republic of Azerbaijan,\\
		H.Javid Avenue 131,  Baku AZ-1143, Azerbaijan\\
	\end{center}
	
	\thispagestyle{empty}
	
	\vskip2cm
	
	%----------------------- abstract ----------------------
	\centerline{\bf ABSTRACT}
	\vskip 4mm

Coupling constants of $\rho$- and $\omega$ - meson-nucleon are related with each other by isospin relation. İn present work, it is aimed at studying the violation (if it exists) of isospin symmetry of the $\omega$-meson  and to study  the temperature dependence of $\omega$-meson-nucleon-delta and $\omega$-meson-delta coupling constants using the soft-wall model of holographic QCD. Firstly, the coupling constants are obtained by applying the profile functions of the vector and fermion fields to the model at a finite temperature. The  interaction Lagrangian is written, including the minimum and magnetic interactions of vector and fermion fields which is defined in the bulk of the 5-dimensional AdS space-time. Secondly, the temperature dependence of the coupling constants $g_{\omega N N}(T)$, $g_{\omega \Delta \Delta}(T)$ and  $g_{\omega N \Delta}(T)$ has been plotted. Comparing $g_{\omega N N}(T)$ with the coupling constant $g_{\rho NN}(T)$, it  obtained that the isospin symmetry of the omega meson is not violated at the finite temperature. It is also observed that the coupling constants of the $\omega$ vector meson with baryons decreases with increasing temperature, and this coupling constant becomes zero near the Hawking phase transition temperature.
	\vspace{2cm}
	%PACS numbers:
	%\today
\end{titlepage}
\section{INTRODUCTION}
There exist several approaches, which play flagship roles in the study of hot hadronic matter. AdS/QCD Anti-de-Sitter/Quantum Chrome Dynamics) approach based on AdS/CFT (Anti-de-Sitter/Conformal field theory) or holographic duality is one such powerful application \cite{1,2,3,4}. The use of this duality allows for solving some particle physics problems, unsolvable within the gauge theory, by employing the gravitational theory;whereas, one of these models is AdS/QCD  models which based on holographic duality. AdS/QCD or holographic QCD has two main models such as hard- and soft-wall models \cite{5,6,7,8}.  A great number of strong interaction problems already have been solved through application of both AdS/QCD models \cite{9,10,11,12,13,14,15,16,17,18,19}. These models are also useful for describing the phase transition of particles as well as the theoretical studies in hot hadronic matter \cite{20,21,22,23,24,25,26,27,28}. 
For instance, the solutions of the equations of motion for meson and fermion fields interacting with the thermal dilaton field at finite temperature are studied in the Refs. \cite{25,26}. Present paper considers the temperature dependence of the $\omega$ meson as continuation of Ref. \cite{35,36}.
Since quantum numbers of iso-singlet $\omega$ and iso-triplet $\rho$ mesons are the same, the main difference between them is the expression of coupling constant of these vector particles with baryons.
Although the temperature dependence of $\omega$ meson nucleon interaction constant was studied in Ref. \cite{39}, present work takes into account the singletness of the omega meson for studying the isospin symmetry of $\omega$ meson. Additionally, it also studies the temperature dependence of  minimal omega meson-delta  $g_{\omega \Delta \Delta}(T)$, and omega meson-nucleon-delta transition coupling constants $g_{\omega N\Delta}(T)$. To investigate violation of  $\omega$ meson isospin symmetry, given  expression $g_{\omega NN}^{(s.w.)}$ / $g_{\rho NN}^{(s.w.)}\approx N_{c}$ has been used.  
 The content of this paper is as follows:
Section II reviews the soft-wall model at finite temperature. Secs. III, V and VI introduce the vector, and baryon fields profil functions in 5-dimensional space-time at finite temperature. 
 In Sec. IV, It has shown the idea of the breaking of chiral symmetry at finite temperature.
 In Sec. VII, It has written the Lagrangian for the vector-spinor interaction in the bulk of AdS utilizing holographic correspondence and has obtained the temperature-dependent integral expression for the $g_{\omega NN}(T)$ at the boundary of QCD.
 In Sec. VIII,  The temperature-dependent integral expressions  $g_{\omega \Delta \Delta}(T)$, and $g_{\omega N\Delta}(T)$ have been obtained at finite temperature.
 In Sec. IX,  The temperature dependence graphs have plotted for $g_{\omega NN}(T)$, $g_{\omega \Delta \Delta}(T)$, and $g_{\omega N\Delta}(T)$.
 Sec.X reviews conclusion.
 
\section{SOFT-WALL MODEL AT FINITE TEMPERATURE}
 In the soft wall model, a dilaton field adding to the model breaks the chiral symmetry with an exponential factor which is made the integral over the $z$ coordinate finite at the IR boundary $(z\rightarrow \infty )$. The parameter k is a scale parameter. 
The form of the dilaton field $\varphi(z)=k^{2}z^{2}$ has been used in the zero temperature case. In thermal soft wall mode, instead of the dilaton field $\varphi(z)$, temperature dependent expansion of dilaton field   $\varphi(r, T)$ is used. Note that for simplicity it is used $r$ instead of $z$ coordinate at finite temperature and 
the relation between tortoise coordinate $r$ and $z$ is as follows:
\begin{equation*}
r=\int \frac{dz}{f(z)}
\end{equation*}
At finite temperature the r coordinate can be
expanded as
\begin{equation}
r\approx z\left[1+\frac{z^{4}}{5z_{H}^{4}}+\frac{z^{8}}{9z_{H}^{8}}\right].
\label{3}
\end{equation}
 Therefore, the action in this model can be written as follows:
\begin{equation}
S=\int d^{4}x dr\sqrt{g}e^{-\varphi(r,T)}L(x,r,T),
\label{1}
\end{equation}
here $ g$ denotes $ g=|det g_{M N}|$ $(M,N=0,1,2,3,5)$ , the extra dimension $r$ varies in the range $0\leqslant  r< \infty$. $ x=(t,\vec x)$ is the set of Minkowski coordinates, $A(r)=log(\frac{R}{r})$, and  $R$ is the AdS space radius. 

For the finite-temperature soft wall model, the temperature dependence of the dilaton field was chosen by multiplying the temperature-dependent function \cite{23} by the $k^2$ parameter \cite{25,26,27}:
\begin{equation}
K^{2}(T)=k^{2}\left[1+\rho(T) \right].
\label{7}
 \end{equation}
 So, $K^{2}(T)$ is the parameter of spontaneous breaking of chiral symmetry and the thermal function $\rho(T)$ up to $T^4$ order has a following form:
\begin{equation} 
		\rho(T)=\frac{9\alpha\pi^{2}}{16}\frac{T^{2}}{12F^{2}}-\frac{ N_{f}^{2}-1}{N_{f}}\frac{T^{2}}{12F^{2}}-\frac{N_{f}^{2}-1}{2N_{f}^{2}}\left(\frac{T^2}{12F^2}\right)^{2} +O\left(T^6\right).
		\label{8}
	\end{equation}
Here  $ \emph F$ is the pseudo-scalar decay constant in the chiral limit and the coefficients $\delta_{T_{1}}$ and  $\delta_{T_{2}}$ are defined by the number of quark flavors $N_f$
  
\begin{equation}
\delta_{T_{1}}=-\frac{ N_{f}^{2}-1}{N_{f}},
\label{9}
 \end{equation}
 and
\begin{equation}
\delta_{T_{2}}=-\frac{N_{f}^{2}-1}{2N_{f}^{2}}. 
\label{10} 
\end{equation}
 
\section{$\omega$ MESON PROFILE FUNCTIONS AT FINITE TEMPERATURE}
Using the meson field equation of motion, the author has derived the $\omega$ meson profile function in the thermal soft wall model of AdS/QCD.
 The vector field $M_{N}(x, r, T)$, which gives the wave function of the $\omega$ meson at the ultraviolet boundary of space-time, consists of $V_L$ and $V_R $ gauge fields, with $M_{N}=1/2\left(V_L+V_R\right)$. The gauge fields belong to the flavor symmetry subgroups $SU(2)_L$ and $SU(2)_R$ respectively, which are part of the flavor group $SU(2)_L\times SU(2)_R$ of the soft wall model of holographic QCD. The action for the vector field is written as follows:
 \begin{eqnarray}
S_M=-\frac{1}{2}\int d^{4}xdr\sqrt{g}e^{-\varphi(r,T)}[\partial_{N}M_{N}(x, r,T)\partial^{N}M^{N}(x, r,T)\nonumber\\
-(\mu^{2}(r,T)+V(r,T))M_{N}(x, r,T)M^{N}(x, r,T)].\label{11}
\end{eqnarray}  
where  $V(r,T)$ is the thermal dilaton potential which reads as
\begin{equation}
V(r,T)=\frac{e^{-2A(r)}}{f^{\frac{3}{5}}(r)}\left[\varphi^{\prime\prime}(r,T)+\varphi^{\prime}(r,T){A}^{\prime}(r)\right], 
\label{12}
\end{equation}
here the prime denotes the $r$ derivative. The temperature dependent 5-dimensional bulk-"mass" $\mu(r,T)$ of the boson field is expressed as
\begin{equation}
\mu^{2}(r,T)=\frac{\mu^{2}}{f^{\frac{3}{5}}(r)}.
\label{13}
\end{equation}
\begin{eqnarray}
 f(r)= 1-\frac{r^4}{r_H^4},
 \label{2}
\end{eqnarray}
where $r_{H}$ is the position of the event horizon and 
  is related to the Hawking temperature as $T=1/(\pi r_{H})$,
 5-dimensional mass $\mu^{2}$, is expressed by the conformal dimension $\Delta =N+L$ of the interpolating operator dual to the meson field. $N$ is the number of partons, and $L=max |L_r|$ is the quark orbital angular momentum for particles. For $\omega$ meson $N=2$, $L=0$, and spin $J=1$. The expression for $\mu^{2}R^{2}$ obtains the simple form \cite{25} as follows:
\begin{equation}
\mu^{2}R^{2}=(\Delta-1)(\Delta-3).
\label{14}
\end{equation} 
 The $M_N$ is Klauza-Klein (KK) modes wave functions corresponding to meson states as follows:
\begin{equation}
M_{\mu}(x,r,T)=\sum_{n}{M_{\mu n}(x)\Phi_{n}(r,T)}.
\label{15}
\end{equation}
 here $\Phi_n (r,T)$ is the temperature depending profile functions.
 EOM (equation of motion) for the $\omega$ meson field will be reduced to the Schröedinger-type equation with the following replacement 
 \begin{equation}
 \phi_{n}(r,T)=e^{-\frac{B_{T}(r)}{2}}\Phi_n (r,T)
 \end{equation}
 with $B_{T}(r)=\varphi(r,T)-A(r)$. In the rest frame of the vector field the EOM gives the following equation for $\phi_{n}(r,T)$ as follows: 
\begin{equation}
\left[-\frac{d^{2}}{dr^{2}}+U(r,T)\right]\phi_{n}(r,T)=M_{n}^{2}(T)\phi_{n}(r,T).
\label{16}
\end{equation}
where $U(r, T)$ is the effective potential and is the sum of both the temperature-dependent and zero-temperature components:
\begin{equation}
U(r,T) =U(r) + \Delta U(r,T).
\label{17}
\end{equation}
 $U(r)$ and $\Delta U(r,T)$ are given as follows:
\begin{equation}
U(r)=k^{4}r^{2}+\frac{(4m^{2}-1)}{4r^{2}},
\label{18}
\end{equation}
\begin{equation}
\Delta U(r,T)=2\rho(T)k^{4}r^{2}.
\label{19}
\end{equation}
where $m=N+L-2$ for $\omega$ meson with two partons, is $m=L$.
The meson mass spectrum $ M_{n}^{2}$ also is the sum of zero and temperature-dependent parts at low temperature as follows:
\begin{equation}
 M_{n}^{2}(T) =\ M_{n}^{2}(0)+\Delta M_{n}^{2}(T),
 \label{20}
\end{equation}
\begin{equation}
\Delta M_{n}^{2}(T)=\rho(T)M_{n}^{2}(0) + \frac{R\pi^{4}T^{4}}{k^{2}},
\label{21}
\end{equation}
\begin{equation}
M_{n}^{2}(0)=4k^2\left(n+\frac{m+1}{2}\right), \\ R =(6n-1)(m+1).
\label{22}
\end{equation}
The solution of equation for the bulk profile $\phi_{n}(r,T)$ of meson field  is given in the following form \cite{25}:
 \begin{equation}
 \phi(r,T)=\sqrt{\frac{2\Gamma(n+1)}{\Gamma(n+m+1)}}K^{m+1}z^{m+\frac{1}{2}}e^{-\frac{K^{2}r^{2}}{2}}L_{n}^{m}(K^{2}r^{2}).
 \label{23}
 \end{equation}
 By taking meson $m=1$, and $n=0$ in ground state for $\omega$ meson, the solution of equation of motion for the bulk profile function  of $\omega$ meson is given the following form \cite{35}:
 \begin{equation}
 M_{0}(r,T)=\sqrt{2}K^{2}r^{\frac{3}{2}}e^{-\frac{K^{2}r^{2}}{2}}L_{0}^{1}(K^{2}r^{2}).
 \label{23-1}
 \end{equation}
\section{CHIRAL SYMMETRY BREAKING AT FINITE TEMPERATURE}
The pseudo-scalar field  $X$, which transforms under the bifundamental representation of $SU(2)_{L}\times SU(2)_{R}$ group, is introduced to perform breaking of the chiral $SU(2)_{L}\times SU(2)_{R}$ symmetry group by using the Higgs mechanism in \cite{29,30,31}. The action for this field has the form as follows:
\begin{equation}
S_{X}=\int d^{4}xdr\sqrt{g}e^{-\varphi(r,T)}Tr\left[|DX|^{2}+3|X|^{2}\right].
\label{24}
\end{equation}
where $D^{M}$ is the covariant derivative and is defined as follows:
\begin{equation}
    D^{M}X = \partial^{M}X - iA^{M}_{L}X+XA^{M}_{R}=\partial^{M}X-i\left[M_M,X\right]-i\{A_M,X\}.
    \label{25}
\end{equation}
Here the last term of Eq. \ref{24} is ignored, since the author only deals with the vector field, so the expected value for this field is \cite{29}:
\begin{equation}
\left< X \right>= \frac{1}{2}am_{q}r+\frac{1}{2a}\Sigma r^{3}=v(r).
\nonumber
\end{equation}
where $m_{q}$ is the mass of $u$ and $d$ quarks, $\Sigma=\left<0|\bar{q}q|0\right> $ is the value of the chiral condensate at zero temperature and $a=\sqrt{N_{c}}/(2\pi)$.
\begin{equation}
\left< X(r,T)  \right> = \frac{1}{2}am_{q}r+\frac{1}{2a}\Sigma (T)r^{3}=v(r,T).
\label{26}
\end{equation}
In \cite{25, 26, 27} it was supposed that the temperature dependence of the $ \Sigma(T)=\left<0|\bar{q}q|0\right>_{T} $ quark condensate is identical to the temperature dependence of the dilaton parameter as  $K^2(T)$ :
\begin{equation}
K^2 (T)=k^2\frac{\Sigma(T)}{\Sigma}.
\label{27}
\end{equation}
The relationship between the quark condensate $\Sigma$ at zero temperature, the number of flavors $N_f$, the condensate parameter $B$, and the pseudoscalar meson decay constant $F$ in the chiral limit is estimated as follows:
\begin{equation*}
\Sigma= - N_{f}BF^{2}
\end{equation*}
Also applies to the finite temperature case:
\begin{equation}
\Sigma(T)= - N_{f}B(T)F^{2}(T).
\label{28}
\end{equation}
Then, according to the Eq.\ref{26}, it can be written as:
\begin{equation}
		\Sigma(T)=\Sigma[1-\frac{N_{f}^{2}-1}{N_{f}}\frac{T^{2}}{12F^{2}}-\frac{N_{f}^{2}-1}{2N_{f}^{2}}(\frac{T^{2}}{12F^{2}})^{2}+O(T^{6})]=\Sigma[1+\Delta_{T}+O(T^{6})], 
		\label{29}
	\end{equation}
 The $ F(T)$ and  $B(T)$ dependencies  were studied in \cite{25}.
\section{SPIN $\frac{1}{2}$ BARYON PROFILE FUNCTION AT FINITE TEMPERATURE}
There are two bulk fermion fields $\left(N_1, N_2\right)$ to describe the two independent chiral components of nucleons [30, 31] at the AdS space-time boundary. Action for the thermal fermion field  $N(x,r,T)$  is written as \cite{26}:
\begin{equation}
S=\int d^{4}x dre^{-\varphi(r,T)}\sqrt{g}{\bar{N}}(x,r,T)D_{\pm }(r)N(x,r,T),
\label{30}
\end{equation}
where the covariant derivative $D_{\pm }(z)$ has the following open form:
\begin{equation}
D_{\pm}(r) =\frac{i}{2}\Gamma^{M}\left[\partial_{M}-\frac{1}{4}\omega_{M}^{ab}\ \left[\Gamma_{a}\Gamma_{b}\right]\right]\mp\left[\mu_F(r, T)+U_{F}(r, T)\right].
\label{31}
\end{equation}
 Here, $\mu_F (r, T)$ is the 5-dimensional mass of the thermal fermion field $N(x,r,T)$ in the soft-wall model of AdS/QCD at finite temperature.
\begin{equation}
\mu_F(r, T)=\mu_F\ f^{\frac{3}{10}}(r, T).
\label{32}
\end{equation}
$\mu_F$ corresponding to the zero-temperature case is defined by the following equation:
\begin{equation}
\mu_F=\emph N_{B}+\emph L-\frac{3}{2},
\label{33}
\end{equation}
where $\emph N_{B}=3$ is the number of partons in the composite fermion and $\emph L$ is the orbital angular momentum ( $\frac{1}{2})$ baryons for $\emph L=0$ spin are considered here). For fermions, the temperature-dependent potential $U_{F}(r, T)$ relates to the zero-temperature cases as follows:
\begin{equation}
U_{F}(r, T)=\varphi (r,T)/f^{\frac{3}{10}}(r, T)
\label{34}
\end{equation}
 The nonzero components of spin connection $ \omega_{M}^{ab} $ are  given by
\begin{equation}
\omega_{M}^{ab}=(\delta_{M}^{a}\delta_{r}^{b}-\delta_{M}^{b}\delta_{r}^{a})\ r f^{\frac{1}{5}}(r, T). 
\label{35}
\end{equation}
The commutator of the Dirac matrices are defined $\sigma^{MN}=\left[\Gamma^{M},\Gamma^N\right]$. These matrices are related to those in the reference frame by the $\Gamma^{M} =e_{a}^{M}\Gamma^{a}$ relation, where
 $e_{a}^{M}=r\times diag \{\frac{1}{f(r)},1,1,1,-f(r)\} $ are the inverse vielbeins and we pass to the reference frame $\Gamma^{a}$ matrices by help of them $ \Gamma^{a}=(\gamma^{\mu },\ -i\gamma ^{5}) $.    
Using $N_{r}(x,r,T)=0$ the author decompose the AdS fermion field into the following left- and right-chirality components
\begin{equation}
N(x,r,T)=N^{R}(x,r,T)+N^{L}(x,r,T),
\label{36}
\end{equation}
which are defined as follows:
$ N^{R}(x,r,T)= \frac{1-\gamma^{5}}{2}N$,   $N^{L}(x,r,T)=\frac{1+\gamma^{5}}{2}N$
with properties
$\gamma^{5}N^{L}=-N^{L}$, $\gamma^{5}N^{R}=N^{R}$.
Kaluza-Klein expansion for the 4-dimensional transverse components of the AdS fields have been written in the terms of the sum of the profile functions $\Phi_{n}^{L, R}(r, T)$
\begin{eqnarray}
N^{R}(x,r,T)=\sum_{n} N_{n}^{R}(x)\Phi_{n}^{R}(r,T),  \nonumber \\
N^{L}(x,r,T)=\sum_{n} N_{n}^{L}(x)\Phi_{n}^{L}(r,T).
\label{37}
\end{eqnarray}
For the nucleons total angular momentum $L$ is $0$ and spin $\emph{J}$ is equal to $\frac{1}{2}$. For this case it is useful to write the  $\Phi_{n}^{L,R}(r,T)$ profiles with the exponential factors
\begin{equation}
\Phi_{n}^{L,R}(r,T)=e^{-\frac{3}{2}A(r)} F_{n}^{L,R}(r,T).
\label{38}   
\end{equation}
 After the substitution of these profile functions into the EOM in the rest frame of nucleon $(\vec{p}=0)$, we get following form of EOM \cite{26} for the $F_{n}^{L,R}(r,T)$ profile functions as follows:
\begin{equation} 
\left[\partial_{r}^2+U_{L,R}(r,T)\right]F_{n}^{L,R}(r,T)=M_{n}^{2}(T)F_{n}^{L,R}(r,T).
\label{39}
\end{equation}
The temperature-dependent spectrum $M_{n}^{2}(T) $ is similar to that at zero-temperature:
\begin{equation}
M_{n}^{2}(T)=4K^2(T)\left(n+m+\frac{1}{2}\right)=4k^2\left(1+\rho (T)\right)\left(n+m+\frac{1}{2}\right).
\label{40}
\end{equation}
$U(r,T)$ in equation (\ref{39}) is the effective potential at finite temperature for the fermion field and can be decomposed into terms dependent on zero- and finite temperature as follows:
\begin{eqnarray}
U_{L/R}(r,T) =U_{L,R}(r)+\Delta U_{L,R}(r,T),
\nonumber \\
\Delta U_{L/R}\left(r,T\right) =2\rho (T)k^{2}\left(k^{2}r^{2}+m \mp \frac{1}{2}\right).
\label{41}
\end{eqnarray}
Here
\begin{equation}
m=N+L-\frac{3}{2}.
\label{42}
\end{equation}
The solutions of the equations \cite{26} are the following profile functions for baryons at finite temperature \cite{26}:
\begin{eqnarray}
F_{L/R}(r,T)=\sqrt{\frac{2\Gamma (n+1)}{\Gamma (n+m_{L}/m_{R}+1)}}K^{m_{L}/m_{R}+1}r^{m_{L}/m_{R}+\frac{1}{2}}e^{-\frac{K^{2}r^{2}}{2}}L_{n}^{m_{L}/m_{R}}\left(K^{2}r^{2}\right), \nonumber\\
\label{43}
\end{eqnarray}
$\Phi(r,T)$ and $F(r,T)$ are the profile functions and obey the normalization conditions.
Where the partonic number$N$ is $N=3$ and the orbital angular momentum $L$ is $L=0$ values for nucleons. In addition,  
$m_{L,R}=m\pm\frac{1}{2}$;therefore, for left and right nucleons $m_{L}=2$ and $m_{R}=1$ respectively.
Solutions of the equation are the following profile functions for  spin $\frac{1}{2}$  baryons (nucleons) at the finite temperatures are as follows \cite{35}:
\begin{equation}
F_{L}(r,T)=\frac{\sqrt2}{\Gamma (3)}K^{2}r^{\frac{5}{2}}e^{-\frac{K^{2}r^{2}}{2}}L_{0}^{2}\left(K^{2}r^{2}\right),
\label{43-1}
\end{equation}
\begin{equation}
F_{R}(r,T)=\sqrt{2}K^{2}r^{\frac{3}{2}}e^{-\frac{K^{2}r^{2}}{2}}L_{0}^{1}\left(K^{2}r^{2}\right),
\label{43-2}
\end{equation}
 
 \section{SPIN $\frac{3}{2}$ BARYON PROFIL FUNCTIONS AT FINITE TEMPERATURE}
 
According to the holographic duality principle, the fields in the bulk of the 5-dimensional AdS space-time which correspond to $\Delta$ baryon operator and the nucleon operator are different at the AdS space-time boundary. Thus, while the Dirac field corresponds to nucleons with spin $\frac{1}{2}$, the Rarita-Schwinger field $\Psi_{M}$ corresponds to the $\Delta$ baryon operators with spin $\frac{3}{2}$ within this space-time.
\cite{20, 21, 22, 23}.
 At finite temperature, the aktion for the Rarita-Schwinger field corresponding to the zero-temperature case can be written as:
\begin{equation}\label{RS}
\int
d^5x\sqrt{G}\left(i\bar{\Psi}_A\Gamma^{ABC}D_B\Psi_{C}-m_1\bar{\Psi}_A\Psi^A-m_2\bar{\Psi}_A\Gamma^{AB}\Psi_B\right)\,,
\end{equation}
where $\Psi_{A}=e_{A}^{M}\Psi_{M}$ and we used notations
$\Gamma^{ABC}=\frac{1}{3!}\Sigma_{\rm{perm}}(-1)^p\Gamma^A\Gamma^B\Gamma^C=\frac{1}{2}(\Gamma^B\Gamma^C\Gamma^A
-\Gamma^A\Gamma^C\Gamma^B)$ and
$\Gamma^{AB}=\frac{1}{2}[\Gamma^A\,,\Gamma^B]$. The Rarita-Schwinger
equations in $AdS_5$  are  written as
\begin{equation}
i\Gamma^A\Big(D_A \Psi_B - D_B \Psi_A\Big) - (m_{1}-m_{2}) \Psi_B + \frac
{(m_{1}+m_{2})}{3}\Gamma_B\Gamma^A\Psi_A = 0
\end{equation}
where $m_1$ and $m_2$ correspond to spinor harmonics on $S^5$ of $\text{AdS}_5 \times S^5$ \cite{24}.

As is known, the Rarita-Schwinger field contains spin 1/2 states of the field in addition to spin 3/2 components. Components with spin 1/2 in 4-dimensional space are eliminated by applying the Lorentz condition to this field:

\begin{equation}
\gamma^{\mu}\Psi_{\mu}=0\,.
\end{equation}

The following Lorentz-covariant constraint will project one of the spin-1/2 components from the Rarita-Schwinger fields into the five-dimensional space corresponding to the four-dimensional space.

\begin{equation}
e_{A}^{M}\Gamma^{A}\Psi_{M}=0\,,
\end{equation}

Then combined with the equations of motion gives $\partial^{M}\Psi_{M}=0$ for a free particle. Also, this field has an additional spin-1/2, $\Psi_{r}$, when descending into four-dimensional space-time at finite temperature. By choosing $\Psi_r = 0$ , the extra spin-1/2 degree of freedom can be reduced, since at finite temperature there should be no extra spinor for the $J=\frac{3}{2}$ baryon field.

\begin{equation}
(iz\Gamma^{A}\Psi_{A} +2i\Gamma^{5} - (m_{1}-m_{2})) \Psi_{\mu}= 0,
\end{equation}
$\quad(\mu=0,1,2,3)$
\begin{equation}
\Psi_{M(R)}=\frac{1}{2}(1+\gamma^{5})\Psi_{N}
\end{equation}
\begin{equation}
\Psi_{M(L)}=\frac{1}{2}(1-\gamma ^{5})\Psi_{M}
\end{equation}
\begin{equation}
[\partial _{r}^{2}-\frac{2(m_{1}-m_{2})+K(T)^{2}r^{2})}{r}\partial _{r}+\frac{2((m_{1}-m_{2})-K(T)^{2}r^{2})}{r^{2}}+p^{2}]\Psi_{L}=0
\end{equation}
\begin{equation}
[\partial _{r}^{2}-\frac{2((m_{1}-m_{2})+K(T)^{2}r^{2})}{r}\partial _{r}+p^{2}]\Psi_{R}=0
\end{equation}
\begin{eqnarray}
f_{L}(r,T)=\sqrt{\frac{2\Gamma (n+1)}{\Gamma (n+m_{L}+1)}}K^{m_{L}+1}r^{m_{L}+\frac{1}{2}}e^{-\frac{K^{2}r^{2}}{2}}L_{n}^{m_{L}}\left(K^{2}r^{2}\right), \nonumber\\
f_{R}(r,T)=\sqrt{\frac{2\Gamma (n+1)}{\Gamma (n+m_{R}+1)}}K^{m_{R}+1}r^{m_{R}+\frac{1}{2}}e^{-\frac{K^{2}r^{2}}{2}}L_{n}^{m_{R}}\left(K^{2}r^{2}\right),
\label{43-3}
\end{eqnarray}
 where
$m_{L,R}=m\pm\frac{1}{2}$.
 The polynomial solutions of these equation are the similar to the spin 1/2 baryon field \cite{36}. 
The profile functions of $\Delta$ baryons $f_{m}(r,T)$ and nucleons $F_{n}(r,T)$ obey normalization conditions as following form:
\begin{equation}
\int_{0}^{\infty}dr e^{-\frac{3}{2}A(r)}f_{m}^{L,R}(r,T)F_{n}^{L,R}(r,T)=\delta_{mn} 
\label{44} 
\end{equation} 

\section{$\omega$ MESON NUCLEON COUPLING CONSTANT AT FINITE TEMPERATURE}

 To obtain the omega meson–nucleon coupling constants in the soft wall model of holographic QCD at finite temperature, first, the interaction Lagrangian between the bulk vector and the nucleon fields is written in the AdS/CFT soft wall model. After some calculation, by using  correspondence between the generating functions of the theories in the 5-dimensional AdS space-time bulk and the 4-dimensional Minkowski space-time boundary is used as following form:
 \begin{equation}
 Z_{AdS}=e^{iS_{int}} = Z_{QCD}.
 \label{46}
 \end{equation}
 By taking variation from the bulk functional $Z_{AdS}$ over the boundary value of the bulk meson field $M_{\mu}^{a}(q)$ nucleon current is founded:
 \begin{equation}
 \left<J_{\mu}^a\right>=-i\frac{\delta Z_{AdS}}{\delta M_{\mu}^{a}(q)}|_{M_{\mu}^{a}=0}.
 \label{47}
 \end{equation}

 There are various types of interactions $L_{int}$ consists of terms that describe these interactions. The first term is the  minimal gauge interaction of the vector field with the current of bulk fermions:

\begin{equation}
L^{(0)}=\bar{N_{1}}e_{A}^{M}\Gamma^{A}M_{M}N_{1}+\bar{N_{2}}e_{A}^{M}\Gamma^{A}M_{M}N_{2}.
\label{48}
\end{equation}

The next terms are related to the 5-dimensional "magnetic" of the bulk spinor field defined by $\Gamma^{MN}$. The 4-dimensional components of this tensor correspond to the magnetic moments of the fermions in the reference frame. The first of these terms is a 5-dimensional generalization of the usual 4-dimensional magnetic interaction as

\begin{eqnarray}
L^{(1)}=ik_{1}e_{A}^{M}e_{B}^{N}\left[\bar{N_{1}}\Gamma^{AB}(F_{L})_{MN}N_{1}-\bar{N_{2}}\Gamma^{AB}(F_{R})_{MN}N_{1}+h.c.\right] \nonumber \\
=ik_{1}e_{A}^{M}e_{B}^{N}\left[\bar{N_{1}}\Gamma^{AB}F_{MN}N_{1}-\bar{N_{2}}\Gamma^{AB}F_{MN}N_{1}+h.c.\right]+ axial \enspace vector \enspace term,
\label{49}
\end{eqnarray}
here $F_{MN}=\partial_MM_N-\partial_NM_M$ is  the field strength tensor of the $M_N$ vector field. 
The $L^{(2)}$ is magnetic moment terms and was constructed in \cite{31} and has a following form:
\begin{eqnarray}
L^{(2)}=\frac{i}{2}k_{2}e_{A}^{M}e_{B}^{N}\left[\bar{N_{1}}X\Gamma^{AB}(F_{R})_{MN}N_{2}+\bar{N_{2}}X^{+}\Gamma^{AB}(F_{L})_{MN}N_{1}-h.c. \right] \nonumber \\ 
=\frac{i}{2}k_{2}e_{A}^{M}e_{B}^{N}\left[\bar{N_{1}}X\Gamma^{AB}F_{MN}N_{2}+\bar{N_{2}}X^{+}\Gamma^{AB}F_{MN}N_{1}+ axial \enspace vector \enspace term \right].
\label{50}
\end{eqnarray}
The interaction Lagrangian term $L^{(2)}$ also includes the interaction with the field $X$ and the this $X$ scalar field changes the chirality of the boundary nucleons and is expressed by the quark condensate $\Sigma$ in the boundary theory. At the limit of QCD theory, the nucleon-$\omega$ meson quark condensate describes the change of chirality of nucleons. So, the general magnetic type Lagrangian has the  following form:
\begin{equation}
L=L^{(1)}+L^{(2)}.
\label{51}
\end{equation}
 Having explicit expressions of the thermal profile functions, one can calculate  the terms of thermal action in the momentum space. This variation gives  the contribution of each Lagrangian term to the nucleon current at finite temperature:
 \begin{equation}
 \left<J_{\mu}\left(p^{\prime},p;T\right)\right>=g_{\omega NN}(r,T)\int dp^{\prime} dp \bar{u}(p^{\prime})\gamma_{\mu}u(p),
 \label{52}
 \end{equation}
The contribution of each Lagrangian term to the coupling constant $g_{\omega NN}(r,T)$  is as shown below.
 $\emph L^{(0)}$  Lagrangian is denoted by $g_{\omega NN}^{(0)nm}(r,T)$ and its integral expression is equal to the following one:
 \begin{eqnarray}
g_{\omega NN}^{(0)nm}(r,T)=3\int_{0}^{\infty }\frac{dr}{r^{4}}e^{-K^{2}r^{2}}M_{0}(r,T)\left[F_{1L}^{*(n)}(r,T)F_{1L}^{(m)}(r,T)\right.\nonumber\\
\left.+F_{2L}^{*(n)}(r,T)F_{2L}^{(m)}(r,T)\right].
\label{53}     
\end{eqnarray}
Between the profile functions of bulk fermion fields, the following relationships are used: $F_{1L}^{(n)} = F_{2R}^{(n)}$, $F_{1R}^{(m)}= -F_{2L }^{(m)}$.

The matrix $\Gamma^{MN}F_{MN}$ is the sum of two types of terms, which are $\Gamma^{5\nu}F_{5\nu}$ and  $\Gamma^{\mu\nu}F_{\mu\nu}$.
  In the  total Lagrangian $L$  the $\Gamma^{5\nu}F_{5\nu}$ terms contribute to the coupling constant $g_{\omega NN}$ and the contribution of this term has following expression at finite temperature:
\begin{eqnarray}
g_{\omega NN}^{(1)nm}(r,T)=-6\int_{0}^{\infty }\frac{dr}{r^{3}}e^{-K^{2}r^{2}}M_{0}^{\prime}(r,T)\left[k_{1}\left(F_{1L}^{*(n)}(r,T)F_{1L}^{(m)}(r,T)\right.\right.\nonumber\\
\left.\left.-F_{2L}^{*(n)}(r,T)F_{2L}^{(m)}(r,T)\right)+k_{2}v(r,T)\left(F_{1L}^{*(n)}(r,T)F_{2L}^{(m)}(r,T)-F_{2L}^{*(n)}(r,T)F_{1L}^{(m)}(r,T)\right)\right].
\label{54}    
\end{eqnarray}

The contribution of the  $\Gamma^{\mu\nu}F_{\mu\nu}$ terms is the following:
\begin{eqnarray}
f_{\omega NN}^{nm}(r,T)=-12m_{N}\int_{0}^{\infty }\frac{dr}{r^{3}}e^{-K^{2}r^{2}}M_{0}(r,T)\left[k_{1}\left(F_{1L}^{*(n)}(r,T)F_{1R}^{(m)}(r,T)-F_{2L}^{*(n)}(r,T)F_{2R}^{(m)}(r,T)\right) \right.\nonumber\\
\left.+k_{2}v(r,T)\left(F_{1L}^{*(n)}(r,T)F_{2R}^{(m)}(r,T)-F_{2L}^{*(n)}(r,T)F_{1R}^{(m)}(r,T)\right)\right]. 
\label{55}   
\end{eqnarray}

Here $m_{N}$ is the mass of the nucleon, $f_{\omega NN}^{nm}(r,T)$  is the contribution of the interaction with the $\omega$ meson due to the nucleon's magnetic moment at finite temperature.
The coupling constant $g_{\omega NN}^{s.w.}(r,T)$ is the sum of two coupling constants at finite temperature:
\begin{equation}
g_{\omega NN}^{s.w.}(r,T)=g_{\omega NN}^{(0)nm}(r,T)+g_{\omega NN}^{(1)nm}(r,T).
\label{56}
\end{equation}
The $g_{\omega NN}^{(0)nm}(r,T)$ coupling constant is interpreted as a "strong charge" and the $f_{\omega NN}^{nm}(r,T)$ constant as a constant of the interaction of the $\omega$ meson with the nucleon using the magnetic moment at finite temperature.
To study the isotopic symmetry of the omega meson at finite temperature, the relation between $\rho$ and $\omega$ meson couplings is used.
\section{the temperature dependence of minimal couplings $g_{\omega \Delta \Delta}(T)$ and  $g_{\omega N \Delta}(T)$}
Considering that $\Delta$ resonance is the dominant decay channel as $\Delta\mapsto N \pi$, $\Delta$  baryons are very close resonance to nucleons. The study of this transition is very important in the study of important nucleon properties such as nucleon potential. For this purpose, after studying the $\Delta$ -nucleon transition form factors \cite{37}, these issues were also considered in the finite temperature medium.  In this work, the author has  investigated the temperature dependence of the omega meson-nucleon-delta transition and  omega meson-delta baryon coupling constants as a continuation of \cite{36} in which meson-nucleon-delta baryon  transition and meson-delta baryon coupling constants were studied at finite temperature.
 The thermal couplings have  expressions of thermal profile functions of the bulk fields. It has been calculated the terms of thermal action in the momentum space by taking the variation from the Lagrangian terms. This variation gives the following contribution of each Lagrangian term to the nucleon current:
The Lagrangian of $\omega$ meson-$\Delta$ baryon interaction is as follows:
\begin{equation}
{\cal L}_{\omega \Delta \Delta} =
\bar\Psi_1^{\nu} \Gamma^{\mu} V_{\mu} \Psi_{1\nu} +\bar\Psi_2^{\nu} \Gamma^{\nu}
V_{\nu} \Psi_{2\nu}
\end{equation}
From corresponding Lagrangian of $\omega$ meson, the $\omega$ meson-$\Delta$ baryon minimal coupling constant $g^{(0)nm}_{\omega \Delta \Delta}(T)$  has following form:
\begin{eqnarray}
g^{(0)nm}_{\omega \Delta \Delta}(T) &=& \! -\int^{\infty}_0 \frac{dz}{r^2}
M_{0}(r,T)\left[\Big(f^{(n)*}_{1L}(r,T) f^{(m)}_{1L}(r,T) + f^{(n)*}_{2L}(r,T)
f^{(m)}_{2L}(r,T)\Big)\right.
\end{eqnarray}

 $M_{0}(r, T)$ is the expression of the
profile function or wave function of $\omega$  meson in the ground state ($n=0$).

The $\omega$ meson nuklon-$\Delta$ baryon transition coupling constant has been obtained from the gauge-
invariant coupling constant of gauge fields with  $\Delta$ resonances and nucleons
 at finite temperature. The Lagrangian for these fields is given by
\begin{eqnarray}\label{Sfnd}
{\cal
L}_{\omega N \Delta}&=&\left[\alpha_1\Big(\bar{\Psi}_1^M\Gamma^N(F_L)_{MN}N_1-\bar{\Psi}_2^M\Gamma^N(F_R)_{MN}N_2\Big)\right. \nonumber\\
\end{eqnarray}
where $\alpha_{1}$ is a parameter \cite{18}. By KK reduction of 5D spinors as $\Psi_{iL,R}(p,r) =
\sum_n f^{(n)}_{iL,R}(p,r) \psi^{(n)}_{L,R} (p)$  for nucleons and
$\Psi_{L,R}(p,r) = \sum_n F^{(n)}_{L,R}(p,r) \psi^{(n)}_{L,R} (p)$ 
for $\Delta$ resonances. The $\omega$ meson-nucleon-$\Delta$ baryon coupling constant at finite temperature can be written similar way to the pion and $\rho$ meson-nucleon-$\Delta$ baryon coupling constants.
\begin{eqnarray}
g^{nm}_{\omega N \Delta}(T) &=& \int^{\infty}_0 dr
\left[\frac{M_{0}(r, T)}{r^2}\Big(\kappa \big(F^{(n)*}_{1L}(r,T)
f^{(m)}_{1R}(r,T) - F^{(n)*}_{2L}(r,T) f^{(m)}_{2R}(r,T)\big)\right].
\end{eqnarray}
\section {numerical results}
The author has studied violation of isospin symmetry of $\omega$ meson  by compering the  $g_{\omega NN}^{(0)nm}(r,T)$ and $g_{\rho NN}^{(0)nm}(r,T)$ coupling constants at finite temperature.

For this reason, since the $g_{\rho NN}(T)$ is already known, the $g_{\omega N N}(T)$ is also studied in this work. In addition,  $g_{\omega \Delta \Delta}(T)$ and $g_{\omega N \Delta}(T)$ is considered by the author.

The author has used MATEMATICA package for numerical calculation and plotted temperature dependence graphs of couplings. The author presented the numerical results for the choice of parameters for two flavor $N_{f}=2$, $F=87$ MeV, three flavor  $N_{f}=3$, $F=100$ MeV. These sets of parameters were taken  from \cite{25}
Additionally, free parameters $k$, $k_{1}$, $k_{2}$,  $m_{q}$ and $\Sigma$ were used for calculate the coupling constants at finite temperature.
The parameters  $k_{1}$ and $k_{2}$  were fixed at the values  $k_{1}=-0.78$ $ GeV^{3}$,  $k_{2}=0.5 $ $GeV^{3}$ in the \cite{31} and $k=383$.  The value of $\Sigma=(368)^{3}$ $MeV^{3}$ and  $m_{q}=0.00145$ $MeV$ parameters were found from the fitting of the $\pi$ meson mass \cite{34}. 

To have an idea of the relative contributions of different terms of Lagrangian, the author present results for the temperature dependencies of the coupling constants separately. In Fig. 1-4 the yellow graph curve represents the $g_{\omega NN}^{(0)nm}(r,T)$ coupling constant, the purple curve shows the $g_{\omega NN}^{(1)nm}(r,T)$ coupling,  the blue shows the $g_{\omega NN}^{(s.w.)}(r,T)$ coupling at finite temperature in the figures below. The green graph curve shows the $f^{nm}_{\omega N N}(r,T)$ coupling constant as well.
In Fig. 3 and Fig.4, the author considers these dependencies for the first excited state ($n=1$) of the nucleons and plot graphs for the different numbers of flavors and observes that all curves in the figures converge at one temperature value. Changing the parameter values does not change the picture in the form and leads to only a slight deformation of the shape in the graph.
Fig. 5 and Fig. 6 show the temperature dependence of the $\omega$ meson $\Delta$ baryon coupling constant and the $\omega$ meson-nucleon- $ \Delta$ baryon coupling constants at different values of the parameters $N_{f}=2$, $F=87$ MEV, $N_{f}=3$, $F=100$ MEV, $N_{f}=4$, $F=130$ MEV, $N_{f}=5$, $F=140$ MEV in ground state ($n=0$) respectively.
\section{conclusion}
The coupling constants of the $\omega$ meson with nucleons, delta baryons, and nucleon-delta baryons transitions within the thermal soft-wall model of holographic QCD have been studied.  The author has plotted each term of the coupling constant as a function of temperature and observed that with increasing temperature the value of these coupling constants decreases and becomes zero at points close to the Hawking temperature. This point corresponds to the confinement-deconfinement transition point, after which the hadrons completely melt and form the quark gluon plasma state. It was found that there is no isospin symmetry breaking of $\omega$ meson at non-zero temperatures

\newpage
\begin{figure}[!ht]
\centering
\includegraphics[scale=0.65]{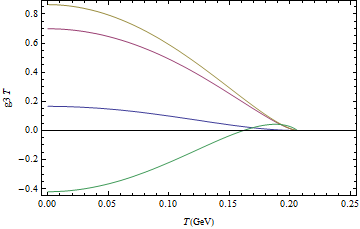}
\caption{Comparison of $g_{\omega NN}^{(0)nm}$, $g_{\omega NN}^{(1)nm}(T)$, $g_{\omega NN}^{(s.w.)}(T)$ and $f^{nm}_{\omega NN}(T)$ coupling constants for the ground state ($n=0$) at the parameter finite  values $N_{f}=2$ and $F=87$ MeV.}
\label{fig:Figure1}
\end{figure}
\begin{figure}[!ht]
\centering
\includegraphics[scale=0.65]{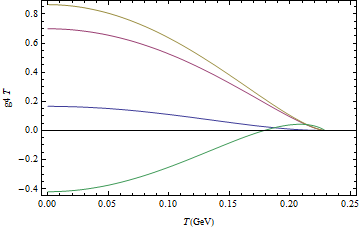}
\caption{Comparison of $g_{\omega NN}^{(0)nm}$,
$g_{\omega NN}^{(1)nm}(T)$, $g_{\omega NN}^{(s.w.)}(T)$ and $f^{nm}_{\omega NN}(T)$ coupling
constants for the ground state ($n=0$) at the parameter values $N_{f}=3$ and $F=100$ MeV }
\end{figure}
\begin{figure}[!ht]
\centering
\includegraphics[scale=0.65]{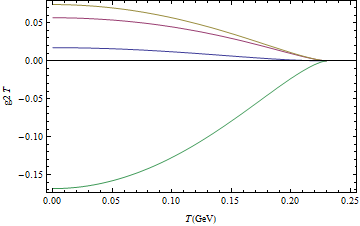}
\caption{ $g_{\omega NN}^{(0)nm}$,
	$g_{\omega NN}^{(1)nm}(T)$, $g_{\omega \Delta \Delta}^{(s.w.)}(T)$ and $f^{nm}_{\omega NN}(T)$ coupling
	constants for the first excited nucleons ($n=1$) at the parameter values $N_{f}=2$ and $F=87$ MeV.}
\label{fig:Figure 2}
\end{figure}
\begin{figure}[!ht]
\centering
\includegraphics[scale=0.65]{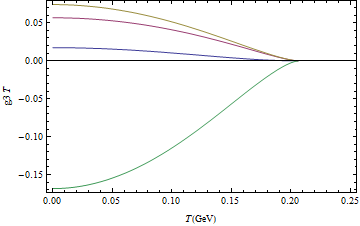}
\caption{$g_{\omega NN}^{(0)nm}$,
	$g_{\omega NN}^{(1)nm}(T)$, $g_{\omega NN}^{(s.w.)}(T)$ and $f^{nm}_{\omega NN}(T)$ coupling
	constants for the first excited nucleons ($n=1$) at the parameter values $N_{f}=3$ and $F=100$ MeV.}
\end{figure}
\begin{figure}[!ht]
\centering
\includegraphics[scale=0.65]{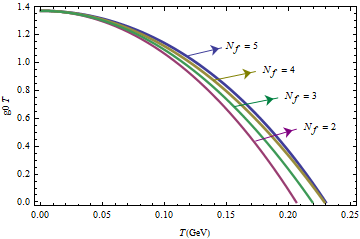}
\caption{The temperature dependence of $g^{0}_{\omega \Delta \Delta}(T)$  for $N_{f}=2$, $F=87$ MEV, $N_{f}=3$, $F=100$ MEV, $N_{f}=4$, $F=130$ MEV, $N_{f}=5$, $F=140$ MEV in ground state ($n=0$).}
\end{figure}
\begin{figure}[!ht]
\centering
\includegraphics[scale=0.65]{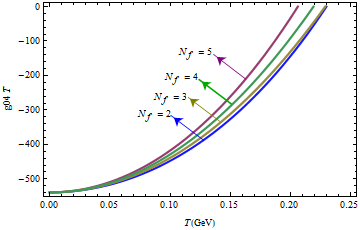}
\caption{The temperature dependence of $g^{0}_{\omega N \Delta}(T)$  for $N_{f}=2$, $F=87$ MEV, $N_{f}=3$, $F=100$ MEV, $N_{f}=4$, $F=130$ MEV, $N_{f}=5$, $F=140$ MEV in ground state ($n=0$).}
\end{figure}

\end{document}